\documentclass[twocolumn]{aastex702}

\usepackage{amsmath}
\usepackage{graphicx}
\usepackage{bm}
\usepackage{booktabs}
\usepackage{apjfonts}
\usepackage{xcolor}
\usepackage{appendix}
\usepackage{url}

\newcommand{\mr}[1]{\mathrm{#1}}

\begin{document}

\title{\Large VLA Observations Confirm AT\,2023mfm as an Off-nuclear Tidal Disruption Event}

\author[0009-0007-3464-417X]{Wenkai Li}
\affiliation{Department of Astronomy, University of Science and Technology of China, Hefei 230026, China}
\affiliation{School of Astronomy and Space Sciences,
University of Science and Technology of China, Hefei 230026, China}
\email[show]{liwenkai@mail.ustc.edu.cn}

\author[0000-0003-0528-202X]{Collin T. Christy}
\affiliation{Steward Observatory and Department of Astronomy, University of Arizona, 933 North Cherry Avenue, Tucson, AZ 85721-0065, USA}
\email[show]{collinchristy@arizona.edu}

\author[0000-0002-8297-2473]{Kate D. Alexander}
\affiliation{Steward Observatory and Department of Astronomy, University of Arizona, 933 North Cherry Avenue, Tucson, AZ 85721-0065, USA}
\email{kdalexander@arizona.edu}

\author[0000-0003-0466-3779]{Itai Sfaradi}
\affiliation{Department of Astronomy, University of California, Berkeley, CA 94720-3411, USA}
\affiliation{Berkeley Center for Multi-messenger Research on Astrophysical Transients and Outreach (Multi-RAPTOR), University of California, Berkeley, CA 94720-3411, USA}
\email{itai.sfaradi@berkeley.edu}

\author[0009-0002-8172-714X]{Xinya Huang}
\affiliation{Department of Astronomy, University of Science and Technology of China, Hefei 230026, China}
\affiliation{School of Astronomy and Space Sciences,
University of Science and Technology of China, Hefei 230026, China}
\email{hxy230663@mail.ustc.edu.cn}

\author[0000-0002-7152-3621]{Ning Jiang}
\affiliation{Department of Astronomy, University of Science and Technology of China, Hefei 230026, China}
\affiliation{School of Astronomy and Space Sciences,
University of Science and Technology of China, Hefei 230026, China}
\email{jnac@ustc.edu.cn}

\author[0000-0003-1792-2338]{Tanmoy Laskar}
\affiliation{Department of Physics \& Astronomy, University of Utah, Salt Lake City, UT 84112, USA}
\email{tanmoy.laskar@utah.edu}

\author{Andrew Mummery}
\affiliation{School of Natural Sciences, Institute for Advanced Study, 1 Einstein Drive, Princeton, NJ 08540, USA}
\email{amummery@ias.edu}

\author[0000-0003-4537-3575]{Noah Franz}
\affiliation{Steward Observatory and Department of Astronomy, University of Arizona, 933 North Cherry Avenue, Tucson, AZ 85721-0065, USA}
\email{nfranz@arizona.edu}

\author[0000-0003-3441-8299]{Adelle J. Goodwin}
\affiliation{International Centre for Radio Astronomy Research, Curtin University, GPO Box U1987, Perth, WA 6845, Australia}
\email{adelle.goodwin@curtin.edu.au}

\author[0000-0001-7946-1034]{Walter W. Golay}
\affiliation{Center for Astrophysics | Harvard \& Smithsonian, 60 Garden St., Cambridge, MA 02138, USA}
\email{wgolay@cfa.harvard.edu}

\author[0000-0003-4768-7586]{Raffaella Margutti}
\affiliation{Department of Astronomy, University of California, Berkeley, CA 94720-3411, USA}
\affiliation{Berkeley Center for Multi-messenger Research on Astrophysical Transients and Outreach (Multi-RAPTOR), University of California, Berkeley, CA 94720-3411, USA}
\affiliation{Department of Physics, University of California, 366 Physics North MC 7300, Berkeley, CA 94720-7300, USA}
\email{rmargutti@berkeley.edu}

\author[0000-0002-7706-5668]{Ryan Chornock}
\affiliation{Department of Astronomy, University of California, Berkeley, CA 94720-3411, USA}
\affiliation{Berkeley Center for Multi-messenger Research on Astrophysical Transients and Outreach (Multi-RAPTOR), University of California, Berkeley, CA 94720-3411, USA}
\email{chornock@berkeley.edu}

\author[0000-0003-3824-9496]{Jiazheng Zhu}
\affiliation{Department of Astronomy, University of Science and Technology of China, Hefei 230026, China}
\affiliation{School of Astronomy and Space Sciences,
University of Science and Technology of China, Hefei 230026, China}
\email{jiazheng@mail.ustc.edu.cn}

\author[0000-0002-3859-8074]{Sjoert van Velzen}
\affiliation{Leiden Observatory, Leiden University, P.O. Box 9513, NL-2300 RA Leiden, The Netherlands}
\email{sjoert@strw.leidenuniv.nl}

\author[0000-0001-7007-6295]{Yvette Cendes}
\affiliation{Department of Physics, University of Oregon, 1371 E 13th Ave, Eugene, OR 97403, USA}
\affiliation{Institute for Fundamental Science, University of Oregon, 1371 E 13th Ave, Eugene, OR 97403, USA}
\email{yncendes@uoregon.edu}

\author[0000-0002-1568-7461]{Wenbin Lu}
\affiliation{Department of Astronomy, University of California, Berkeley, CA 94720-3411, USA}
\affiliation{Theoretical Astrophysics Center, University of California, Berkeley, CA 94720-3411, USA}
\email{wenbinlu@berkeley.edu}

\author[0009-0000-9210-8964]{Jimmy Lynch}
\affiliation{Department of Physics, University of Oregon, 1371 E 13th Ave, Eugene, OR 97403, USA}
\affiliation{Institute for Fundamental Science, University of Oregon, 1371 E 13th Ave, Eugene, OR 97403, USA}
\email{jlyn@uoregon.edu}

\begin{abstract}

We report new radio observations of the tidal disruption event (TDE) AT\,2023mfm, which we identified as a high-confidence candidate in a systematic search for off-nuclear TDEs. High-resolution NSF Karl G. Jansky Very Large Array C-band (6 GHz) imaging resolves two radio sources: one consistent with the host-galaxy nucleus and one offset by $0.651\pm 0.036''$ ($1.06\pm0.06$ kpc), consistent with the Zwicky Transient Facility and Pan-STARRS1 positions of AT\,2023mfm. These observations confirm the off-nuclear nature of AT\,2023mfm, demonstrating the power of high-resolution radio imaging to validate off-nuclear TDE candidates and reveal hidden off-nuclear massive black holes.

\end{abstract}


\section{Introduction}

Most massive galaxies host nuclear massive black holes (MBHs), but some MBHs are predicted to reside offset from their host-galaxy nuclei through galaxy mergers, gravitational-wave recoil after MBH binary mergers, or three-body interactions in MBH triples~\citep{Yao2025}. Two-body relaxation among stars bound to the MBH, or close encounters with unbound stars as the MBH moves through the host, can bring stars within the MBH's tidal radius and produce off-nuclear tidal disruption events (TDEs).

Off-nuclear TDEs are a rapidly developing field and have been discovered in X-rays (e.g., 3XMM J215022.4-055108:~\citealt{Lin2018}; EP240222a:~\citealt{Jin2025}) and the optical (e.g., AT\,2024tvd:~\citealt{Yao2025}, radio-confirmed as off-nuclear by~\citealt{Sfaradi2025}; AT\,2025abcr:~\citealt{Stein2026}). These events demonstrate that TDEs are powerful probes of hidden off-nuclear MBHs, linking individual transients to galaxy--MBH assembly histories.

\begin{figure*}
\centering
\includegraphics[width=\textwidth]{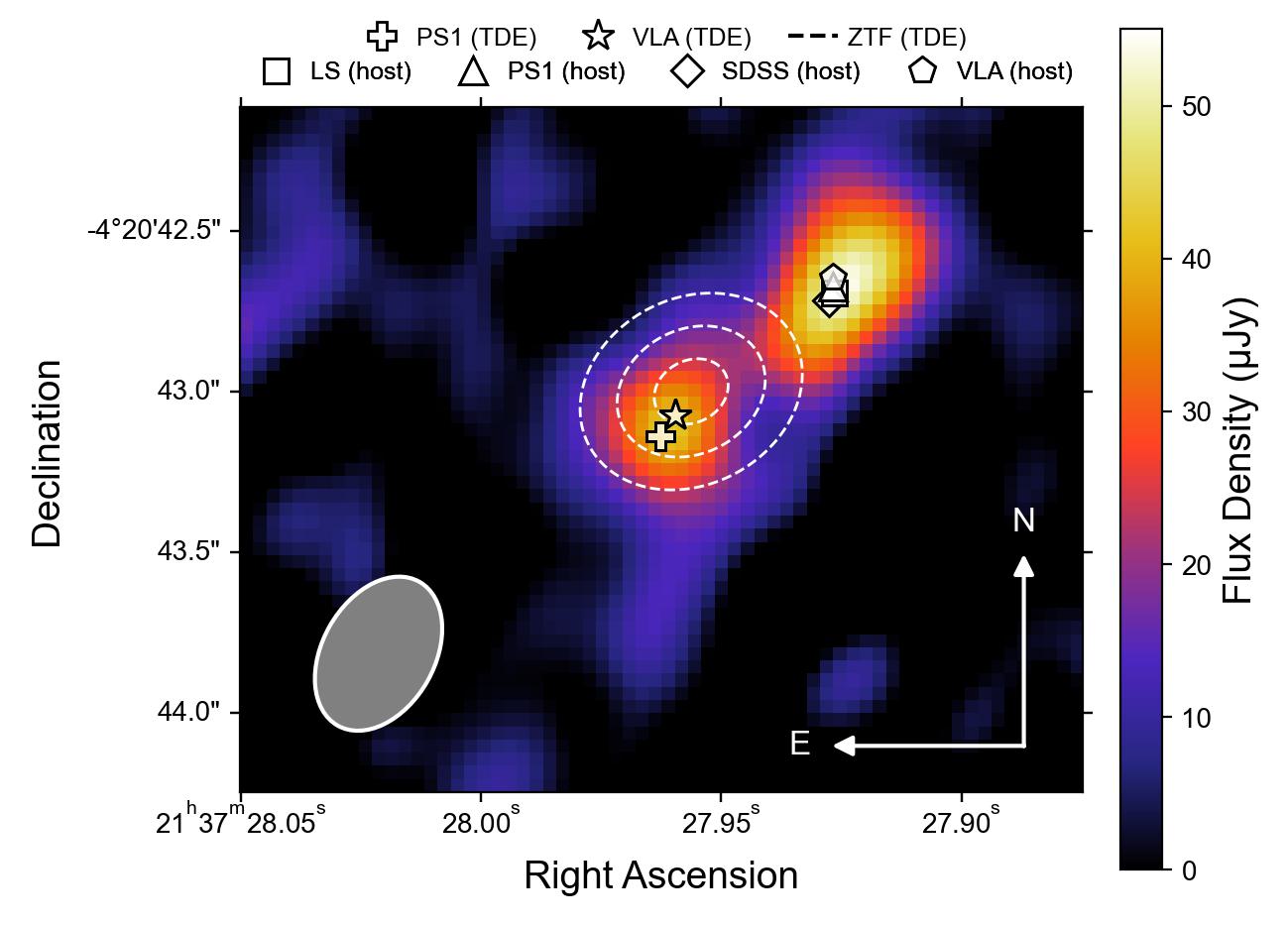} 
\caption{VLA A-configuration C-band (6 GHz) image of AT\,2023mfm on 2026 April 11, revealing distinct radio emission at the TDE location (SE) and host-galaxy nucleus (NW). The white cross marks the PS1/TNS TDE position; white ellipses show the $1\sigma$--$3\sigma$ ZTF contours; the diamond/triangle/square mark the SDSS/PS1/LS host-galaxy centroids, respectively; the star/pentagon mark the VLA detections. The gray ellipse shows the synthesized-beam FWHM.}
\label{fig:VLA}
\end{figure*}

Here we report NSF Karl G. Jansky Very Large Array (VLA) observations confirming AT\,2023mfm as off-nuclear. We assume $H_0 = 70~\mr{km~s^{-1}~Mpc^{-1}}$, $\Omega_{\mr{m}} = 0.3$, and $\Omega_{\Lambda} = 0.7$. Dates are in UT, coordinates are J2000, and uncertainties are $1\sigma$ unless otherwise stated.


\section{Observations and Astrometry} \label{sec_obs_ana}

AT\,2023mfm was discovered by the Zwicky Transient Facility (ZTF) on 2023 June 28~\citep{Fremling2023}. Its Transient Name Server (TNS) position was later updated to $\alpha=21^{\rm h}37^{\rm m}27.961^{\rm s}$, $\delta=-04^\circ20'43.16''$ using Pan-STARRS1 (PS1) data.\footnote{\url{https://www.wis-tns.org/object/2023mfm}} It was spectroscopically classified as a TDE based on a broad H$\alpha$ feature, with host nebular lines at a consistent redshift of $z=0.087$~\citep{Chornock2023}.

We identified AT\,2023mfm as a high-confidence off-nuclear TDE candidate in a systematic astrometric analysis of reported TDEs. Details of our sample selection, astrometry, and error analysis will be presented in Li et al. (2026, in prep.). Briefly, we measured transient positions and compared them with archival host-galaxy centroids from the Sloan Digital Sky Survey (SDSS), PS1, and the DESI Legacy Imaging Surveys (LS). Sources whose host-galaxy centroids lay outside the $3\sigma$ transient-position ellipse and were separated from the transient by more than the instrument-specific empirical astrometric threshold were flagged as strong off-nuclear candidates. Applying this procedure to ZTF data for AT\,2023mfm gives $\alpha=21^{\rm h}37^{\rm m}27.956^{\rm s}$, $\delta=-04^\circ20'43.03''$, with a $3\sigma$ ellipse of $0.301''\times0.238''$, offset from the SDSS/PS1/LS host-galaxy centroids by $0.545''$, $0.580''$, and $0.571''$, respectively. The independent PS1/TNS position agrees with the ZTF position, further supporting this off-nuclear interpretation. A similar conclusion was independently reached by \citet{Stein2026}.

We serendipitously observed AT\,2023mfm with the VLA in A-configuration on 2026 April 11 (1018 days post-discovery) under program \texttt{26A-167} (PI: Christy), which tracks the long-term radio evolution of TDEs. The data were reduced in the Common Astronomy Software Applications package (CASA; \citealt{CASATeam2022}) and imaged with the CASA task \texttt{tclean}. The C-band (6 GHz) image resolves two radio point sources (Figure~\ref{fig:VLA}). We measured positions and flux densities with \texttt{imfit}, fitting an elliptical Gaussian fixed to the synthesized-beam size. The southeastern and northwestern sources in Figure~\ref{fig:VLA} have flux densities of $S_{\nu,\rm SE}=\mathrm{46}\pm\mathrm{9}~\mu{\rm Jy}$ and $S_{\nu,\rm NW}=\mathrm{60}\pm\mathrm{6}~\mu{\rm Jy}$, corresponding to luminosities of $\nu L_{\nu,\rm SE}=(5.2\pm1.0) \times 10^{37}~{\rm erg~s^{-1}}$ and $\nu L_{\nu,\rm NW}=(6.8\pm0.7)\times 10^{37}~{\rm erg~s^{-1}}$. For absolute positional uncertainties, we conservatively added in quadrature a $0.0515''$ systematic term, $10\%$ of the synthesized-beam major-axis FWHM, to both sky-coordinate directions.\footnote{\url{https://science.nrao.edu/facilities/vla/docs/manuals/oss/performance/positional-accuracy}} With this error budget, their positions are $(\alpha_{\rm SE},\delta_{\rm SE})=(21^{\rm h}37^{\rm m}27.9581^{\rm s} \pm 0.0038^{\rm s},-04^\circ20'43.092''\pm0.063'')$ and
$(\alpha_{\rm NW},\delta_{\rm NW})=(21^{\rm h}37^{\rm m}27.9253^{\rm s} \pm 0.0036^{\rm s},-04^\circ20'42.665''\pm0.057'')$. They are separated by $0.651\pm 0.036''$ ($1.06\pm0.06$ kpc at $z=0.087$), where the uncertainty reflects the relative positional error and does not include the absolute astrometric systematic. The southeastern source is consistent with the ZTF/PS1 TDE position, while the northwestern source is consistent with the host-galaxy nucleus approximated by the SDSS/PS1/LS centroids. We also detect the northwestern source in X-band (10 GHz) and Ku-band (15 GHz), and tentatively detect the southeastern source in Ku-band at $\gtrsim3\sigma$, with $23\pm7~\mu{\rm Jy}$. In L-band (1.5 GHz) and S-band (3 GHz), we detect only one point source, likely a blend of the two sources because of the coarser angular resolution.

To assess possible contamination by an unrelated nearby transient, we performed two checks. First, because the ALeRCE ZTF light curve~\citep{Forster2021} shows a late-time optical bump, we split the ZTF data at MJD $=60196$ and measured pre- and post-bump positions, finding only a $0.057''$ difference, consistent with zero within the empirical astrometric uncertainties and supporting a common off-nuclear origin. Second, a $5''$ search around the off-nuclear radio source found no additional ZTF alert or TNS-reported transient, disfavoring association with a different transient.

Together, the positional agreement of the radio sources with the optical TDE position and host-galaxy nucleus, respectively, and the absence of another nearby transient confirm that the off-nuclear radio source is associated with AT\,2023mfm. The radio luminosity and optically thin spectrum of the off-nuclear source are comparable to those of TDEs at $\sim1000$ days post-discovery~\citep{Cendes2024}.


\section{Discussion and Summary} \label{sec_dis_con}

AT\,2023mfm's classification spectrum shows broad H$\alpha$ and nebular lines at a consistent redshift~\citep{Chornock2023}, disfavoring a line-of-sight coincidence with the identified host. We estimate a host-galaxy stellar mass of \(\sim 10^{10.7}-10^{11.0}\,M_\odot\) from the SDSS \(i\)-band luminosity and \(g-i\) color, and independently from the WISE \(W1\) luminosity and \(W1-W2\) color~\citep{Zibetti2009,Cluver2014}, consistent with the massive, old hosts of previously reported off-nuclear TDEs (e.g.,~\citealt{Lin2018};~\citealt{Jin2025};~\citealt{Yao2025};~\citealt{Stein2026}) and lying near the stellar-mass range where the volumetric density of off-nuclear MBHs is predicted to peak~\citep{Guolo2026}. Archival LS imaging shows possible dwarf/satellite-like galaxies and low-surface-brightness diffuse structure southwest of the host, suggesting a rich, possibly ongoing tidal/merger history, consistent with displaced-MBH formation channels.

Among these channels, we favor a minor-merger origin: AT\,2023mfm could have occurred around an infalling satellite's MBH before it sank to the galaxy center, or around an MBH displaced by a possible subsequent three-body interaction. Two points support this. First, the nuclear source has a C--Ku-band spectral index of $0.25\pm0.13$ in our VLA epoch, consistent with compact AGN-like emission. AT\,2023mfm's off-nuclear location therefore suggests an additional displaced MBH in the same galaxy, disfavoring recoil of the central black hole itself. Second, AT\,2023mfm's relatively low \(g\)-band peak luminosity of \(\sim2\times10^{43}~{\rm erg~s^{-1}}\) is more consistent with a lower black hole mass than with the black hole mass expected from the host-galaxy mass~\citep{Mummery2024,Mummery2026}.

In summary, we identify AT\,2023mfm as a high-confidence off-nuclear TDE candidate from ZTF/PS1 astrometry in our systematic analysis of reported TDEs. We confirm this result with high-resolution VLA imaging, resolving two sources separated by $0.651\pm 0.036''$ ($1.06\pm0.06$ kpc): one consistent with the host-galaxy nucleus and one with the optical TDE position. This radio confirmation places AT\,2023mfm among the small but growing sample of off-nuclear TDEs, demonstrating the power of high-resolution radio imaging to validate such candidates and reveal hidden off-nuclear MBHs.

\begin{acknowledgments}
The National Radio Astronomy Observatory and Green Bank Observatory are facilities of the U.S. National Science Foundation operated under cooperative agreement by Associated Universities, Inc.
\end{acknowledgments}

\bibliography{ref}{}
\bibliographystyle{aasjournal}

\end{document}